\documentclass[baaa]{baaa}

\usepackage[pdftex]{hyperref}
\usepackage{subfigure}
\usepackage{natbib}
\usepackage{helvet,soul}
\usepackage[font=small]{caption}

\contriblanguage{1}

\contribtype{2}

\thematicarea{7}

\title{Diffuse radio emission from merger shocks\\ in simulated galaxy clusters}

\titlerunning{Diffuse radio emission from merging galaxy clusters}

\author{
S.E.~Nuza\inst{1,2}
}

\authorrunning{S.E.~Nuza}

\contact{snuza@iafe.uba.ar}

\institute{
Instituto de Astronom\'ia y F\'isica del Espacio, CONICET--UBA, Argentina
\and
Consejo Nacional de Investigaciones Cient\'ificas y T\'ecnicas, Argentina
}

\resumen{Los cúmulos galácticos son las estructuras gravitacionalmente ligadas más grandes del Universo. Como tales, durante eventos de fusión cumular con sistemas análogos, liberan una cantidad enorme de energía que es disipada a través de la formación de ondas de choque y turbulencia en el medio intracúmulo (ICM), siendo este último el plasma caliente que permea su interior. Los choques suelen ser sitios ideales para la aceleración de electrones, los cuales, en presencia de campos magnéticos en el ICM, son capaces de generar las estrucuras elongadas no térmicas típicamente observadas en la periferia de cúmulos galácticos dinámicamente perturbados que se conocen con el nombre de {\it radio relics}. En esta contribución analizamos una re-simulación hidrodinámica de cúmulos galácticos en colisión, extraída de un gran conjunto de simulaciones cosmológicas {\it zoom-in} pertenecientes al {\sc Three Hundred Project}, con el fin de estudiar la evolución y diversidad de los choques de fusión y su emisión difusa en radio asociada dentro del marco del escenario de aceleración difusiva.}

\abstract{Galaxy clusters are the largest gravitationally-bound structures in the Universe. As such, during merger events with similar systems, they release an enormous amount of energy that is dissipated through the formation of shock waves and turbulence in the intracluster medium (ICM), the hot ionised plasma permeating the cluster volume. These shock waves are believed to be ideal sites for electron acceleration that, in the presence of ubiquitous magnetic fields in the ICM, are capable of producing elongated non-thermal radio features typically observed in the outskirts of dynamically disturbed clusters, also known as {\it radio relics}. In this work, we analyse a hydrodynamical re-simulation of merging galaxy clusters, extracted from a large set of {\it zoom-in} cosmological simulations of {\sc The Three Hundred Project}, to study the evolution and diversity of merger shocks and their associated diffuse radio emission within the framework of the diffusive shock acceleration scenario.}

\keywords{
galaxies: clusters: intracluster medium --- shock waves --- radiation mechanisms: non-thermal}

\begin{document}

\maketitle

\section{Introduction}
\label{intro}

Diffuse radio emission from galaxy clusters has been preferentially observed in systems undergoing strong dynamical disruptions \citep{Cassano10}. The origin of such emission, not linked to any individual radio source, is believed to be the result of cosmic ray (CR) (re)acceleration within the intracluster medium (ICM). During cluster merger events, an enormous amount of energy is released to the surrounding medium that is later dissipated through the formation of shock waves and turbulence in the ICM. Of special interest is the formation of elongated non-thermal radio structures seen in the outskirts of colliding galaxy clusters known as {\it radio relics}.   

Several mechanisms have been proposed to explain the spectral and morphological properties of radio relics. The currently accepted scenario favours a link between the {\it merger shocks} produced during galaxy cluster collisions and radio relics. Merger shocks are thought to be ideal sites for electron (re)acceleration, plausibly through diffusive shock acceleration (DSA) or a similar mechanism, being the specific details currently unknown \citep[see e.g.][and references therein]{Nuza17}. According to this scenario, a population of aging CR electrons (CRe) are responsible for the observed non-thermal emission in the radio band which, in the presence of magnetic fields in the ICM, are able to produce synchrotron radiation.  

Cosmological simulations have been used in the past to model the formation and diversity of radio relics. For instance, in \cite{Nuza17}, we used a state-of-the-art sample of re-simulated galaxy clusters, including radiative gas physics, star formation and feedback, to study the morphological properties of merger shocks and the origin of observed radio relic correlations. 

In this work, we take advantage of a galaxy cluster simulation drawn from a novel set of re-simulated galaxy cluster regions extracted from the dark matter (DM) MultiDark cosmological simulation. These galaxy cluster re-simulations comprise the so-called {\sc Three Hundred Project} \citep{Cui18} and are stored with a larger temporal sampling than previous numerical efforts, allowing us to better follow shock evolution during cluster mergers.

\begin{figure*}[!t]
\centering
\includegraphics[width=\columnwidth]{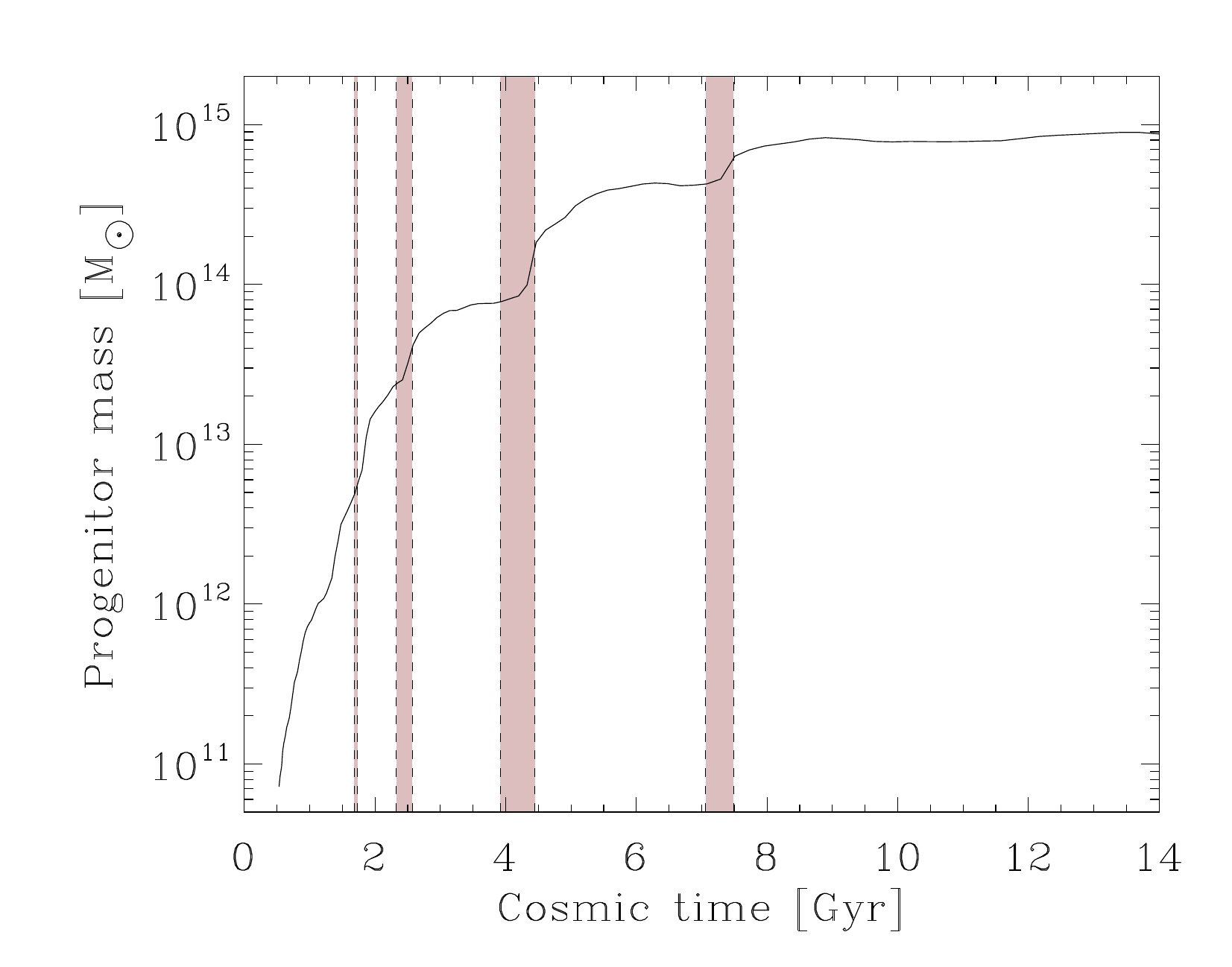}\hspace{0.1cm}\includegraphics[width=\columnwidth]{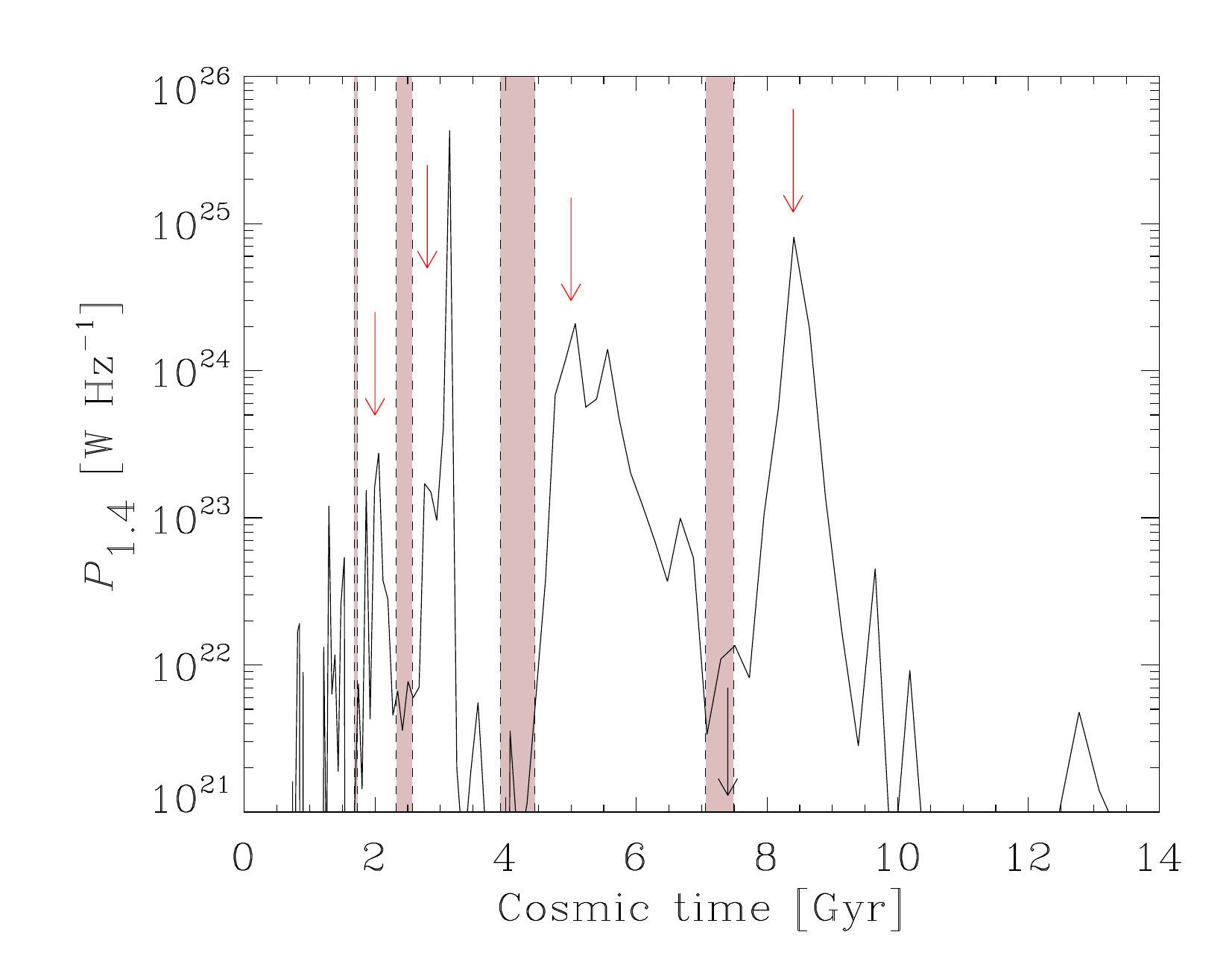}
\caption{\emph{Left-hand panel:} Mass of the most massive progenitor in the re-simulated galaxy cluster region vs. cosmic time. \emph{Right-hand panel:} Radio power per unit frequency at $1.4\,$GHz across the entire lifespan of the cluster. Red arrows highlight radio luminosity peaks. In both panels, major mergers with $\Delta M/M\geq0.5$ are shown by the vertical red bands. The black arrow indicates the core-passage time of the last merger.}
\label{fig1}
\end{figure*}

This proceeding is organised as follows. In Sec.~\ref{simu} we present the main characteristics of the galaxy cluster simulation used in this work. In Sec.~\ref{radio_model} we briefly describe the main assumptions of our non-thermal radio emission model and the identification of structure formation shocks. In Sec.~\ref{results} we discuss the mass assembly and radio emission lightcurve of the simulated cluster across cosmic time as well as some implications for shock evolution during cluster mergers. Finally, in Sec.~\ref{summary} we summarise our main results.

\section{The simulated galaxy cluster}
\label{simu}

In this work, we use a simulated galaxy cluster region performed within the context of the {\sc Three Hundred Project} \citep{Cui18}, a suite of 324 spherical {\it zoom-in} re-simulations of galaxy cluster regions of radius $15\,h^{-1}\,$Mpc extracted from the $1\,h^{-3}\,$Gpc$^3$ DM-only MDPL2
MultiDark simulation \citep{Klypin16}. The cosmological model adopted is consistent with the Planck 2015 cosmology \citep{Planck16}. Galaxy cluster regions were resimulated with the {\sc Gadget-X} code \citep[e.g.][]{Beck16} including full hydrodynamics, metal-dependent gas cooling, an UV background radiation field, and other relevant sub-grid astrophysical processes such as star formation, feedback from supernovae/active galactic nuclei and black hole growth. Mass resolution at the beginning of each re-simulation is $1.27\times10^9\,h^{-1}\,$M$_{\odot}$ and $2.36\times10^8\,h^{-1}\,$M$_{\odot}$, for DM and gas particles, respectively. At $z=0$, our selected cluster region hosts a central cluster of mass $5.83\times10^{14}\,h^{-1}\,$M$_{\odot}$. 

\section{The non-thermal radio emission model}
\label{radio_model}

The amount of radio emission produced in structure formation shocks is estimated in the same way as \cite{Nuza17}. The main assumption of the model is that a fixed fraction of available electrons get accelerated at the shock fronts acquiring an energy distribution consistent with DSA to be later advected by the downstream plasma, where CRe suffer from radiative looses. In the presence of magnetic fields, CRe age, loosing energy via synchrotron emission. Additionally, inverse Compton (IC) collisions with cosmic microwave background (CMB) photons also play a role. The aged CRe energy spectrum at time $t$ after injection is taken from \cite{Kardashev62}
\begin{align*}
    N(\mathcal{E},t) = A\,\mathcal{E}^{-s}\left[1-\left(\mathcal{E}_{\rm max}^{-1}+B t\right)\mathcal{E}\right]^{s-2}{\rm ,}
    \nonumber
\end{align*}
\noindent
where $A$ is a local normalization constant, $\mathcal{E}$ is the electron energy normalized to its rest mass, $s$ is the slope of the DSA injection spectrum, $\mathcal{E}_{\rm max}$ is the maximum normalized energy of accelerated electrons and $B$ is a constant accounting for synchrotron and IC looses. By integrating the CRe energy spectrum across the shock front it is possible to compute the total non-thermal emission per unit frequency and shock area in the downstream region as a function of magnetic field and local gas properties. Magnetic fields are assumed to scale with local electron density \citep{Dolag01}, producing simulated profiles consistent with observations of the Coma cluster \citep[][]{Bonafede10}. 

Synthetic shock fronts are selected by searching for gas particles fulfilling a set of criteria such as convergent flows and entropy/density jumps. Finally, a conservative estimate for the Mach number is computed using the Rankine-Hugoniot conditions \citep{Landau59}. Further details concerning the radio emission model and shock identification in the simulations can be found in \cite{Nuza12,Nuza17} and references therein.

\begin{figure}[!t]
\centering
\includegraphics[width=\columnwidth]{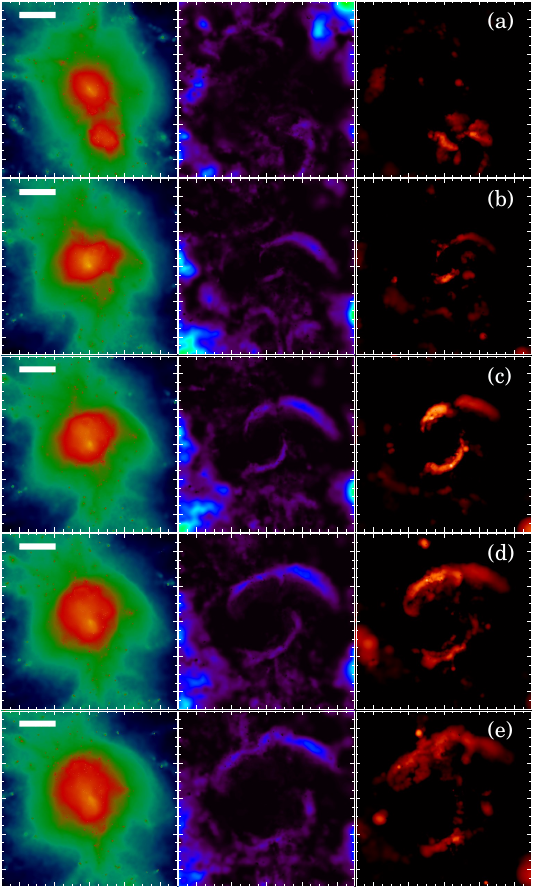}
\caption{Snapshots of the last major binary merger in our re-simulated cluster region for the projected gas density (left column), Mach number (middle column) and radio power (right column) at cosmic times with respect to the core-passage (i.e., $t_{\rm cp}=0$) of (a) $t=-633\,$Myr, (b) $t=212\,$Myr, (c) $t=439\,$Myr, (d) $t=670\,$Myr, (e) $t=904\,$Myr. To guide the eye, solid bars indicate a linear (comoving) scale of $2\,$Mpc. Simulated merger shock Mach numbers after core-passage are $\mathcal{M}\sim2-4$. Radio luminosity magnitudes in the right column can be read directly from the right-hand panel of Fig.~\ref{fig1}.}
\label{fig2}
\end{figure}

\section{Results}
\label{results}

\subsection{Mass assembly and radio lightcurve}

The left-hand panel of Fig.~\ref{fig1} shows the cluster progenitor mass as a function of cosmic time. Clearly, there are four significant merger events (shown by the red-shaded areas) during the cluster formation history. In this region, all significant mergers happen relatively early in the history of the universe, including the last massive merger, which comprises two sub-clusters of masses 
$3.05\times10^{14}\,h^{-1}\,$M$_{\odot}$ and $1.26\times10^{14}\,h^{-1}\,$M$_{\odot}$ mutually approaching in a fairly radial orbit.

The radio luminosity lightcurve at $1.4\,$GHz for all radiation within $R_{200}$ of the progenitor is shown in the right-hand panel of Fig.~\ref{fig1}. This figure clearly demonstrates that every merger event is followed by a peak in radio luminosity, as expected from the formation of merger shocks. The overall lightcurve is generally very noisy as smaller shocks appear and disappear within the cluster volume as a result of minor accreted substructures, typical of a cosmological environment. Interestingly, radio luminosity generally peaks after mergers with a time delay as high as $\sim1\,$Gyr (see below). In close to head-on collisions, this is owing to the time it takes for axially-propagating shocks to form, from launch time until they are fully illuminated by non-thermal radiation, and eventually fade away.   

\subsection{Merger shocks in a cosmological environment}

The evolution of cluster merger shocks within a realistic cosmological context can be seen more easily in Fig.~\ref{fig2}. This figure shows different stages of the most massive binary merger starting at a cosmic time of around $7\,$Gyr in Fig.\ref{fig1}. Firstly, while sub-clusters are approaching, the shock distribution is mainly the result of minor accreted substructures and gas towards the densest regions of the re-simulated volume. Before core-passage, less energetic shock waves are launched in various directions owing to perturbations in the ICM. These shocks, however, are usually subdominant and, therefore, underluminous in comparison with axial counterparts (i.e., in a direction roughly coincident with the merger axis). After core-passage the onset of axial shocks is clearly seen in the second and third rows with shock fronts expanding towards the outskirts of the merged galaxy cluster. As time elapses, the shocks peak in luminosity, and the diffuse emission eventually fades out with decreasing density, as shown by the right-hand panel of Fig.~\ref{fig1}.       

\section{Summary}
\label{summary}

In this work, we used a re-simulation of a galaxy cluster region from {\sc The Three Hundred Project} to model the formation of radio relics by following the evolution of shocks formed during galaxy cluster collisions. Typical simulated relic structures appear after core-passage peaking from several Myr up to $\sim 1\,$Gyr later. Merger shocks are preferentially formed along the merger axis where relics are usually more energetic. Our simulation shows that radio shocks resembling observations are a natural outcome of cluster mergers in a cosmological environment. However, a suite of simulated galaxy clusters is necessary to generate larger synthetic shock samples that can be compared with relic observations at different evolutionary stages.

\begin{acknowledgement}
The author acknowledges the support provided by UBACyT 20020170100129BA.
\end{acknowledgement}


\bibliographystyle{baaa}
\small
\bibliography{bibliografia}
 
\end{document}